\RequirePackage{lineno}
\documentclass[pre,
twocolumn,
eqsecnum,
amsmath,amssymb,
amsmath,amssymb, aps,
]{revtex4}

\usepackage{epsfig}
\usepackage{amsfonts}
\usepackage{amsmath}
\usepackage{amssymb}
\usepackage{bm} 
\usepackage[framemethod=PStricks]{mdframed}
\usepackage{lipsum}
\usepackage{float}
\usepackage{color}
\usepackage{pdflscape}
\usepackage{graphicx}
\usepackage{todonotes}

\newcommand{\be}{\begin{equation}}
\newcommand{\ee}{\end{equation}} 
\newcommand{\bea}{\begin{eqnarray}} 
\newcommand{\eea}{\end{eqnarray}}

\begin{document}

\title{Onset of Intermittency in Stochastic Burgers Hydrodynamics}

\author{G.B. Apolin\'ario$^1$, L. Moriconi$^1$, and R.M. Pereira$^2$}
\affiliation{$^1$Instituto de F\'\i sica, Universidade Federal do Rio de Janeiro, \\
C.P. 68528, CEP: 21945-970, Rio de Janeiro, RJ, Brazil}
\affiliation{$^2$Laborat\'orio de F\'isica Te\'orica e Computacional, Departamento de F\'\i sica, Universidade Federal de Pernambuco,
50670-901 Recife, PE, Brazil}


\begin{abstract}
We study the onset of intermittency in stochastic Burgers hydrodynamics, as characterized by the statistical behavior of negative velocity gradient fluctuations. The analysis is based on the response functional formalism, where specific velocity configurations - the {\it{viscous instantons}} - are assumed to play a dominant role in modeling the left tails of velocity gradient probability distribution functions. We find, \textcolor{black}{as expected on general grounds}, that the field theoretical approach becomes meaningful in practice only if the effects of fluctuations around instantons are taken into account. Working with a systematic cumulant expansion, it turns out that the integration of fluctuations yields, in leading perturbative order, to an effective description of the Burgers stochastic dynamics given by the renormalization of its associated heat kernel propagator and the external force-force correlation function.
\end{abstract}


\maketitle


\section{Introduction}

Burgers one-dimensional hydrodynamics, introduced long ago \cite{burgers1,burgers2} as a simpler model 
designed to illustrate some aspects of Navier-Stokes turbulence \cite{frisch}, 
has been, as actually foreseen by von Neumann at the dawn of the computational era \cite{von_neumann}, 
a valuable testing ground for the development of alternative approaches and new ideas in the 
framework of the statistical theory of turbulence \cite{bec_khanin}.

It is worth emphasizing that the Burgers model is more than just a mathematical toy. The
Burgers equation has been applied to realistic problems in the fields of nonlinear acoustics \cite{gurba_mala}, 
cosmology \cite{zeldovich, gurba_saichev}, critical interface growth \cite{kpz}, traffic 
flow dynamics \cite{musha_higuchi,chow_etal}, and biological invasion \cite{petro_bai}.

A number of theoretical efforts have been devoted to the study of intermittent fluctuations of fluid
dynamic observables, such as velocity gradients $\xi \equiv \partial_x u(x,t)$, or velocity differences, 
$\delta_x u \equiv u(x,t) - u(-x,t)$, in statistically homogeneous and stationary states of
the stochastic version of the Burgers model \cite{polyakov,
gura-migdal, khanin_etal, balko_etal, bec, chernykh-stepanov, mori, grafke_etal, friedrich_etal}.
Positive fluctuations of $\xi$ or $\delta_x u$, related to spatially smooth velocity field configurations, 
are sub-Gaussian random variables \cite{polyakov,gura-migdal}. In contrast, the presence
of velocity shocks in Burgers dynamics leads to extremely intermittent negative fluctuations of these observables, which 
can be described, in principle, by fat-tailed probability distribution functions \cite{khanin_etal, balko_etal, bec, chernykh-stepanov, mori, grafke_etal}, still the matter of current research.

An important point was made in the analytical study put forward in Ref. \cite{balko_etal}, where specific velocity field configurations -- the so-called viscous instantons -- were conjectured to be the dominant structures for a statistical account of large negative fluctuations of $\xi$. It follows that the left tail of the velocity gradient probability distribution functions (vgPDFs), which can be written, without loss of generality, as $\rho_g(\xi) = \exp [ - S(\xi) ]$, should have its asymptotic behavior given by $S(\xi) \sim |\xi|^\frac{3}{2}$, a result later validated by numerical evaluations of the viscous instanton solutions by Chernykh and Stepanov \cite{chernykh-stepanov}. However, as it has been noted in the remarkable numerical {\it{tour de force}} by Grafke {\it{et al}}. \cite{grafke_etal}, even though the asymptotic form of $S(\xi)$ is presently far beyond the reach of direct numerical simulations, the instanton computational strategy is still able to give reasonable answers for the local stretching exponent $\theta(\xi) \equiv d \ln S(\xi)/d \ln (|\xi|)$. The same authors have found, furthermore, that a satisfactory matching between the vgPDF tails obtained from numerical studies and the ones provided by the instanton configurations can be achieved only if the random force strength parameter is multiplied by a Reynolds number dependent adjustment factor. A detailed \textcolor{black}{analytical} investigation of why such an empirical ``noise renormalization procedure'' works is the central aim of our work.

We apply, in the following discussion, field theoretical techniques formerly introduced in the analytical approach to vgPDFs in Lagrangian turbulence \cite{mori2,apol}, where it was found, similarly, that renormalizations of the heat-kernel propagator and of the force-force correlation function play a fundamental role in the description of the vgPDFs' tails. The essential idea of the method consists in the integration, by means of a specifically designed cumulant expansion, of arbitrary fluctuations around the instanton solutions, derived from the Martin-Siggia-Rose-Janssen-de Dominicis (MSRJD) response functional formalism \cite{msr, dominicis, janssen, cardy}. \textcolor{black}{It is natural to expect that corrections to the instanton evaluations of vgPDFs' tails have to be supplemented, for the sake of accuracy, by subdominant fluctuation contributions. As a matter of fact, extensive numerical studies of fluctuations in the instanton approach to Burgers turbulence have been established only very recently through the application of importance sampling and hybrid Monte Carlo techniques \cite{margazoglou_etal, ebener_etal}.}

This paper is organized as follows. In the next section, we describe the specific details of our path-integral approach to the improved derivation of vgPDF tails, which relies on the perturbative integration of fluctuations around instanton solutions, within the cumulant expansion framework. In Sec. III, we discuss the transition from the low to the high Reynolds number regime, related to the crossover between the weak to strong coupling domains in the field theory context. We then show how our results and the empirical ones by Grafke {\it{et al}}. \cite{grafke_etal} come together into a consistent theoretical picture. In Sec. IV, we determine the range of validity of the perturbative treatment, which breaks down at strong coupling. Finally, in Sec. V, we summarize our findings and point out directions for further research.

\section{Field Theoretical Setup}

To start our analysis, we write down the evolution equation for the velocity field, $u = u(x, t)$, in the stochastic Burgers model. In dimensionless form it is given as \cite{grafke_etal}
\be
u_t + u u_x = u_{xx} + g \phi \ , \  \label{burgers_eq}
\ee 
where $\phi = \phi(x,t)$ is a zero-mean Gaussian random field used to model large-scale forcing, with correlator
\be
\langle \phi(x,t) \phi (x',t') \rangle = \chi(x-x') \delta (t -t') \ , \  \label{phi-c}
\ee
which is peaked at wavenumber $k=0$ and broadened in Fourier space within a region of size $\Delta k \sim 1$.
In other words, $L \equiv 1$ ($ \sim \Delta k^{-1}$) is taken to be the random force correlation length, defined as the largest relevant length scale in the flow. Note that the intensity of forcing is given by the noise strength parameter $g$. While most of our considerations in this section are general, we will eventually adopt, as it has been addressed in former works \cite{balko_etal, grafke_etal, gotoh-kraichnan},
\be 
\chi(x) = (1-x^2) \exp \left (  - \frac{x^2}{2} \right ) = - \partial_ x^2 
\exp \left ( - \frac{x^2}{2} \right ) \ , \ 
\ee 
a case study of particular interest, due to its simple formulation and good analytical properties. Furthermore, once the viscosity $\nu$ and the integral length scale $L$ are normalized to unit in Eq. (\ref{burgers_eq}), by defining the Reynolds number as $Re = L^{4/3}\sqrt[3]{\langle (\partial_x u)^2 \rangle / \nu^2}$ we get $Re = \sqrt[3]{g^2/2}$ \cite{comment1}. 

The vgPDFs can be computed in the MSRJD formalism as path-integrations over the velocity field 
$u(x,t)$ and its conjugate auxiliary field $p(x,t)$, combined with an ordinary integration over a Lagrange multiplier 
variable $\lambda$ as 
\bea
&&\rho_g(\xi)  = \langle \delta(u_x|_0 - \xi) \rangle \nonumber \\
&&= {\cal{N}}^{-1} \int Dp Du \int_{- \infty}^{ \infty} d \lambda \exp \{ -S[u,p,\lambda ; g] \} \ , \
\label{vgPDF}
\eea
where $\cal{N}$ is an unimportant normalization constant (to be supressed from now on, in order to simplify notation), $u_x|_0$ is the velocity gradient taken at $(x,t) = (0,0)$, and $S[u,p,\lambda ; g]$ is denoted as the MSRJD action,
\bea
&&S[u,p,\lambda ; g] = \frac{g^2}{2} \int d t d x ~ p (\chi * p ) + \nonumber \\
&&+i \int d t d x ~ p(u_t+u u_x - u_{xx}) - i\lambda (u_x |_0 - \xi ) \ , \ \label{msr_action}
\eea
with $\chi * p \equiv \int dx'\chi(x-x') p(x',t)$.

The saddle-point method is a standard tool to find the asymptotic form of vgPDF tails, provided that they decay 
faster than $\exp(-c|\xi|)$ for any arbitrary $c>0$, as it is actually observed from numerical studies of Burgers turbulence \cite{gotoh-kraichnan}. The saddle-point configurations $u^c$, $p^c$, and $\lambda^c$ that extremize the MSRJD action are named instantons \cite{gura-migdal, falko_etal}. In our specific problem, they can be obtained as the solutions of the Euler-Lagrange variational equations
\be
 \left. \frac{\delta S}{\delta u}\right\vert_{u^c,p^c,\lambda^c} \!\!\!\!\!\!\! =0
 \ \ \ , \ \ \ 
 \left.\frac{\delta S}{\delta p}\right\vert_{u^c,p^c, \lambda^c} \!\!\!\!\!\!\! =0 
  \ \ \mbox{and} \ \ 
 \left.\frac{\partial S}{\partial \lambda }\right\vert_{u^c,p^c, \lambda^c} \!\!\!\!\!\!\! =0   \ . \ \label{sp_eqs}
\ee
It is convenient to rescale $p(x,t)$ and $\lambda$ as
\begin{equation}
    p \rightarrow \frac{p}{g^2} \ \ \mbox{and} \ \
    \lambda \rightarrow \frac{\lambda}{g^2} \mbox{,} \label{rescaling}
\end{equation}
so that the MSRJD action in (\ref{vgPDF}) is rescaled as
\be 
S[u,p,\lambda;g] \rightarrow \frac{1}{g^2} S[u, p, \lambda; 1]  \label{s_to_s}
\ee
and the Euler-Lagrange equations stated in (\ref{sp_eqs}) become
\bea
 &&u_t + u u_x - u_{xx} = i\chi * p \ , \ \label{sp1} \\
 &&p_t+ u p_x + p_{xx} = \lambda \delta(t) \delta'(x) \label{sp2} \ , \ \\
 &&\xi = u_x|_0 \label{sp3} \ . \ 
\eea 
As we see from (\ref{s_to_s}), the noise strength $g$ has been factored out from the expression for the action, a simple observation that will be of great importance later on in our arguments. It is clear, furthermore, from the above equations, that the saddle-point solutions $p^c(x,t)$ and $\lambda^c$, if existent, are pure imaginary numbers, since we look for real velocity instantons $u^c(x,t)$.

\textcolor{black}{When dealing with instantons, one needs, in general, to worry about the existence of degenerate families of saddle-point solutions, associated to symmetries of the action, like translation or gauge invariance. The Fadeev-Popov method is the usual procedure to eliminate such redundant solutions \cite{mori, coleman}. However, in the formalism addressed here, we bypass the degeneracy issue through the explicit assignment of the spacetime point $(x,t)=(0,0)$ as the symmetry center around which the instantons evolve (Eqs. (\ref{sp1}) and (\ref{sp2}) are, in fact, not translationally invariant).}

Eqs. (\ref{sp1}) and (\ref{sp2}) have to be solved forward and backward in time, respectively, in the time domain 
$- \infty < t \leq 0$, with $u(x, - \infty) = p (x, - \infty) = 0$, and the additional boundary conditions given by Eq. (\ref{sp3}) and $p(x,0^+) = 0$ \cite{gura-migdal} (equivalent to $p(x,0^-) = -\lambda \delta'(x)$, which amounts, 
in Fourier space, to $\tilde p^c(k,0^-) = - i \lambda k$).

Chernykh and Stepanov have proposed a fruitful self-consistent numerical strategy to solve the above saddle-point equations \cite{chernykh-stepanov}. One neglects, from the start, the boundary condition (\ref{sp3}), trading it, as a counterpart, for an arbitrary fixed value of $\lambda$. The Chernykh-Stepanov method establishes a sequence of progressively better approximations to the exact numerical instantons,
\be
\{ u_1(x,t) \equiv 0, u_2(x,t), u_3(x,t), \ ... \ \}
\ee
and
\be
\{ p_1(x,t), p_2(x,t), p_3(x,t), \ ... \  \}
\ee
which, up to specific optimization strategies \cite{grafke_etal2} is generated as follows: at the $n_{th}$ iteration step, substitute $u(x,t)$ in Eq. (\ref{sp2})
by $u_n (x,t)$ to find $p_n(x,t)$. The field $p_n(x,t)$ is, then, substituted in Eq. (\ref{sp1}), which is solved to yield the velocity field $u_{n+1}(x,t)$. If this procedure converges, typically in $L^2$ norm, iterations can be carried out until $u(x,t)$ and $p(x,t)$ are obtained up to some desired accuracy. The velocity gradient $\xi$ is defined, {\it{a posteriori}}, from Eq. (\ref{sp3}), with the help of the last iterated velocity field. It turns out, from extensive computational analyses, that $\xi$ is a monotonically increasing function of $\lambda$, and that the left asymptotic vgPDF tails can be numerically addressed, in principle, along the lines of the instanton approach \cite{grafke_etal}. It is important to note, however, that the Chernykh-Stepanov method may require further numerical tricks to attain convergence for $|\lambda|$ large enough.

Once $u^c(x,t)$, $p^c(x,t)$ and $\lambda^c$ are available, we perform the following substitution in the path integral expression (\ref{vgPDF}) (of course, after the mappings (\ref{rescaling}) and (\ref{s_to_s}) have already been implemented),
\bea 
&&u(x,t) \rightarrow u^c(x,t) + u(x,t) \ , \ \label{subs_u} \\ 
&&p(x,t) \rightarrow p^c(x,t) + p(x,t) \ , \ \label{subs_p} \\
&&\lambda \rightarrow \lambda^c + \lambda \ . \ \label{subs_lambda}
\eea
We have introduced, in the RHS of Eqs. (\ref{subs_u}-\ref{subs_lambda}), the fluctuations $u(x,t)$, $p(x,t)$ and $\lambda$, around their respective saddle-point solutions. The MSRJD action is rewritten, accordingly, as
\bea
&&S[u,p, \lambda; 1]  \rightarrow S[u^c + u,p^c + p, \lambda^c + \lambda; 1] \nonumber \\
&&\equiv S_c[u^c,p^c] + S_0[u,p] + S_1[u^c,u,p^c,\lambda] \ , \
\eea
where $S_c$, $S_0$, and $S_1$ are, respectively, the saddle-point action, the sum of all quadratic forms in the $u$ and $p$ fields that do not depend on $p^c$ and $u^c$, and finally, $S_1$ is the contribution that collects all the terms that have not been included in $S_c$ and $S_0$. We have
\bea 
&&S_c  \equiv \frac{1}{2} \int d t d x  ~ p^c (\chi * p^c ) + \nonumber \\
&&+ i \int d t d x ~ p^c(u^c_t+u^c u^c_x - u^c_{xx})  \nonumber \\
&&= ({\hbox{using Eq. (\ref{sp1})}}) = - \frac{1}{2} \int d t d x  \ p^c (\chi * p^c)  \ , \ \nonumber \\
\label{sp_action} \\
&&S_0  =  \int d t d x  \left \{ \frac{1}{2}\ p (\chi * p)  + i \ p (u_t - u_{xx}) \right \} \ , \ \nonumber \\
\eea 
and, up to second order in the fluctuating fields,
\be 
S_1 =  i \int d t d x \left\{ p^c u u_x - p_x u^c u  \right\} 
- i \lambda u_x|_0 \ . \ 
\ee 
It is clear that $S_c$ is a functional of the instanton fields, which on their turn depend on the velocity gradient $\xi \equiv u^c_x |_0 $. Hence we can write, more synthetically, that $S_c = S_c(\xi)$.
\begin{figure}[ht]
\hspace{0.0cm} \includegraphics[width=.2\textwidth]{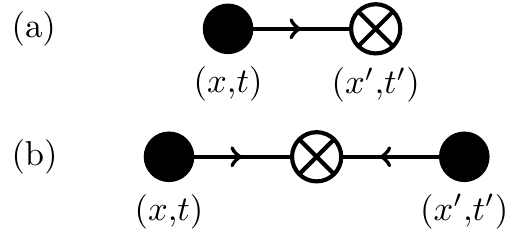}
\caption{Feynman diagrammatic representation of (a) the heat kernel propagator $\langle u(x,t) p(x',t') \rangle_0$ and (b) the velocity-velocity correlator $\langle u(x,t) u(x',t') \rangle_0$.}
\label{propagators}
\end{figure}
\begin{figure}[ht]
\hspace{0.1cm} \includegraphics[width=.46\textwidth]{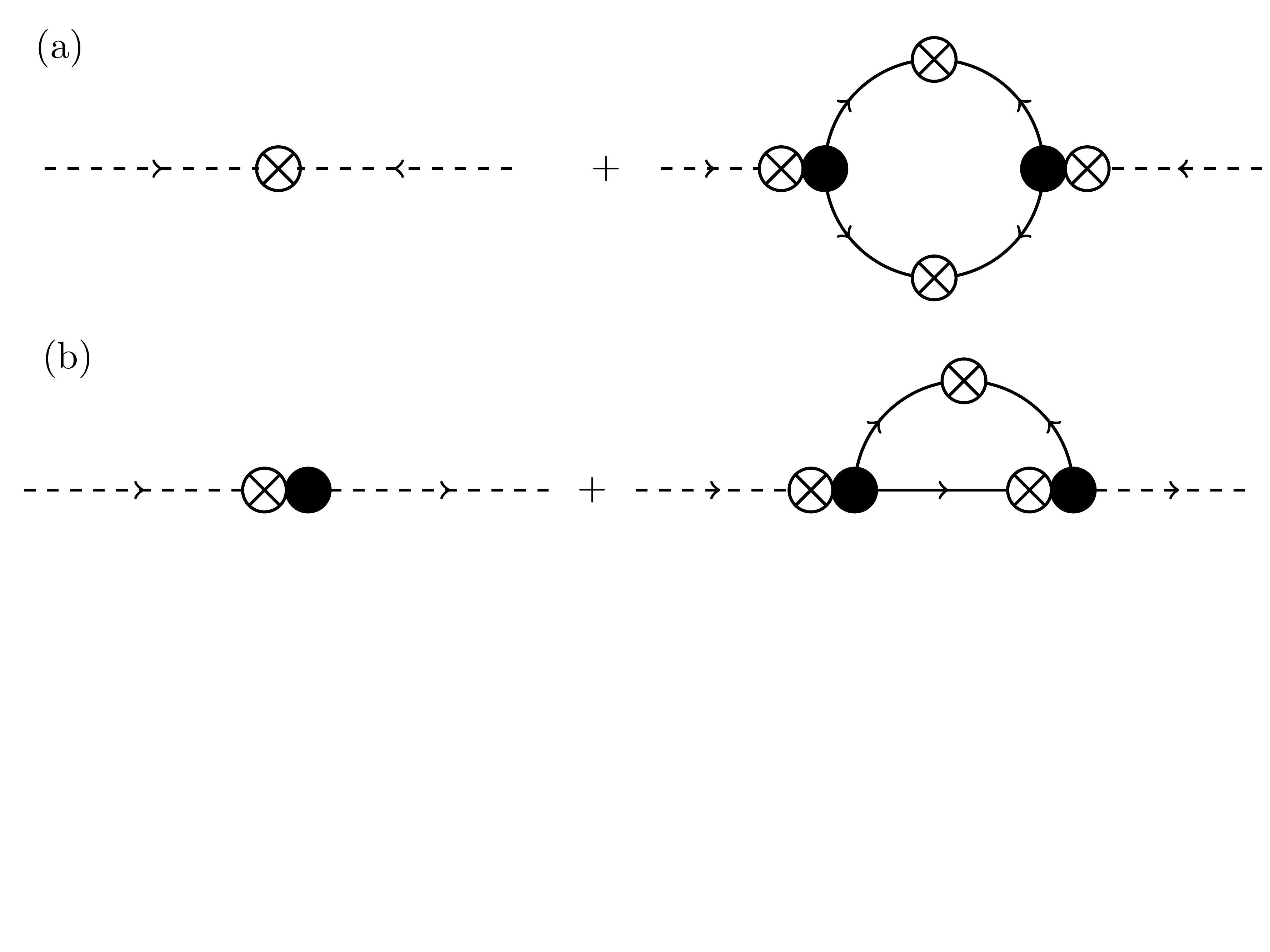}
\vspace{-2.5cm}

\caption{One loop contributions for the renormalization of (a) the noise kernel and (b) the heat kernel propagator in the effective MSRJD action.
Incoming and outgoing dashed lines are associated to the instanton fields $p^c(x,t)$ and $u^c(x,t)$, respectively, as they appear in Eqs. (\ref{I1}) and
(\ref{I2}).}
\label{oneloopcorrections}
\end{figure}
Now, taking into account the instanton solutions, we can reformulate the vgPDF, Eq. (\ref{vgPDF}), as
\bea 
&&\rho_g(\xi) = \exp \left ( - \frac{1}{g^2} S_c(\xi) \right ) \times \nonumber \\
&&\times \int Dp Du 
\int_{- \infty}^\infty d \lambda 
\exp \left [ - \frac{1}{g^2} ( S_0 + S_1 ) \right ] \nonumber \\
&& \propto \exp \left ( -\frac{1}{g^2} 
S_c(\xi) \right ) 
\int_{- \infty}^\infty d \lambda \left \langle \exp \left ( - \frac{1}{g^2} 
S_1 \right )
\right \rangle_0 \ , \  \nonumber \\ \label{vgpdf2}
\eea
where $\langle(...) \rangle_0$ stands for expectation values computed in the linear stochastic model defined by the MSRJD action $S_0$. 

Having in mind perturbative developments in cases where the fluctuation-dependent contributions are small relative to the leading saddle-point results, that is,
\be 
S_c(\xi) \gg g^2 \left | \ln \left [ \int_{- \infty}^\infty d \lambda \left \langle \exp \left ( - \frac{1}{g^2} 
S_1 \right ) 
\right \rangle_0 \right ] \right | \ , \
\ee
we can resort to the cumulant expansion method for evaluating (\ref{vgpdf2}). We obtain, considering contributions up to second order in the instanton fields and $\lambda$,
\bea
&&\left \langle \exp \left ( - \frac{1}{g^2} S_1 \right )
\right \rangle_0 = \nonumber \\
&& = \exp \left \{ - \frac{1}{g^2} \langle S_1 \rangle_0  +   \frac{1}{2 g^4} 
\left [ \langle (S_1)^2 \rangle_0 - \langle S_1  \rangle_0^2 \right ]  \right \} \ . \ \nonumber \\
\label{c-expansion}
\eea
Adding the terms between curly brackets in (\ref{c-expansion}) to $-S_c(\xi)/g^2$ we get, by definition, $ - \Gamma / g^2$, where $\Gamma$ is referred to as the {\it{effective MSRJD action}}, i.e.,
\be
\Gamma \equiv S_c + \langle S_1 \rangle_0  -   \frac{1}{2 g^2} 
\left [ \langle (S_1)^2 \rangle_0 - \langle S_1  \rangle_0^2 \right ] \ . \
\ee
\textcolor{black}{The perturbative integration of fluctuations around the saddle-point solutions by means of cumulants is in fact a standard approximation in field theory, as already discussed long ago, for instance, in Ref. \cite{langouche_etal}.}

The basic building blocks needed to evaluate (\ref{c-expansion}) are the correlation functions
\bea
&& G_{pu}(x,x',t,t') \equiv \langle p(x,t) u(x',t') \rangle_0 = \nonumber \\
&&= -\frac{i g^2}{2 \sqrt{\pi (t'-t)} }
\exp \left [ - \frac{(x-x')^2}{4 (t'-t)} \right ] \Theta(t'-t) \nonumber \\ 
\label{Gpu}
\eea 
and
\bea 
&& G_{uu}(x,x',t,t') \equiv \langle u(x,t) u(x',t') \rangle_0 = \nonumber \\
&&= \frac{g^2}{2 \sqrt{1+2|t-t'|}} \exp \left [ - \frac{(x-x')^2}{2(1+ 2|t-t'|)} \right ] 
\ , \ \nonumber \\ 
\label{Guu}
\eea
which are graphically identified to the Feynman diagrams depicted in Fig. \ref{propagators}.

It is not difficult to show, from (\ref{Gpu}) and (\ref{Guu}), that 
$\langle S_1 \rangle_0 = 0$ and
\be 
\langle (S_1)^2 \rangle_0 = I_1 [p^c,u^c] + I_2 [p^c] - \lambda^2 \langle (u_x|_0)^2 \rangle_0 \ , \ \label{S1squared}
\ee 
with
\bea 
&&I_1 [p^c,u^c] \equiv  \nonumber \\
&&\equiv \int_{t,t'<0} dt dt'dx dx'~ p^c(x,t) u^c(x',t') H_1(x,x',t,t')   \ , \ \nonumber \\
\label{I1} \\
&&I_2 [p^c] \equiv \nonumber \\
&&\equiv \int_{t,t'<0} dt dt'dx dx'~ p^c(x,t) p^c(x',t') H_2(x,x',t,t') \ , \ \nonumber \\
\label{I2}
\eea 
where
\bea 
&&H_1(x,x',t,t') = \nonumber \\
&&= -2 \partial_x [ G_{uu}(x,x',t,t') \partial_x G_{pu}(x',x,t',t) ] \ , \   \\
&&H_2(x,x',t,t') = \frac{1}{2} \partial_x^2 [G_{uu}(x,x',t,t')]^2 \ . \ 
\eea
Note that $I_1[p^c,u^c]$ and $I_2[p^c]$, both of $\mathcal{O} (g^4)$, are, in diagrammatic representation, 
the one-loop contributions which renormalize, respectively, the heat and the noise kernels associated to the 
original stochastic Burgers equation (\ref{burgers_eq}). See Fig. \ref{oneloopcorrections}.

The overall effect of perturbative contributions can always be conventionally accounted by a redefinition of the noise strength parameter $g$ in the expression for the vgPDF, $\rho_g( \xi) \propto \exp ( - S_c(\xi) /g^2 )$, obtained at leading order, as given in Eq. (\ref{vgpdf2}). As a matter of fact, we are led to a particularly simple formulation in the present context. Define the $g$-independent coefficient
\be 
c(\xi) \equiv - \frac{I_1 [p^c,u^c] + I_2 [p^c]}{2g^4 S_c } \ . \ \label{c-xi}
\ee
Using (\ref{vgpdf2}), (\ref{c-expansion}), and integrating over $\lambda$ in the Gaussian approximation given by (\ref{S1squared}), we get, from (\ref{c-xi}),
\be 
\rho_g(\xi) \propto \exp \left ( -  \frac{S_c (\xi) }{g_R^2} \right ) \ , \ \label{eff_vgpdf}
\ee 
where
\be 
g_R \equiv \frac{g}{ \sqrt{1 + c(\xi)g^2}} \label{eff_noise}
\ee 
defines an {\it{effective noise strength}} parameter, which is, in principle, a velocity-gradient dependent quantity that encodes the effects of fluctuations around the instantons, up to the lowest non-trivial order in the cumulant perturbative expansion.

\section{The Onset of Intermittency}

Eq. (\ref{eff_noise}) suggests, in fact, a simple criterion for the consistency of the perturbative analysis. 
It is indicated, from that result, that the cumulant expansion is meaningful, up to second order, if $|c(\xi)|g^2$ is reasonably smaller than unity. It follows, immediately, that for any fixed velocity gradient $\xi$, the cumulant expansion will break down for $g$ large enough. Similarly, since  (as we will see) $c(\xi)$ is a positive monotonically increasing function of $|\xi|$, the cumulant expansion framework becomes inadequate for large enough $|\xi|$ at any fixed $g$. 

The consideration of strong coupling regimes implied by $g \gg 1$ (the ones which have high Reynolds numbers) and/or asymptotically large velocity gradient fluctuations is, thus, precluded from the cumulant expansion approach. The perturbative analysis, nevertheless, is actually useful to model the shape of vgPDF left tails in the non-Gaussian region, where $|\xi| > g$, for not very large $g$. We expect, on physical grounds, that as the noise strength $g$ grows and incipient turbulent fluctuations associated to flow instabilities come into play, the onset of non-Gaussian behavior gets captured by dominant instanton contributions ``dressed" by cumulant corrections.

It is important, before proceeding, to comment on the challenging technical difficulties associated to the evaluations of $S_c(\xi)$,  $I_1 [p^c,u^c]$, and $I_2 [p^c]$, given, respectively, by Eqs. (\ref{sp_action}), (\ref{I1}) and (\ref{I2}), the essential ingredients in the derivation of vgPDF tails. It turns out that the associated integrations based on the numerical instantons are extremely demanding in terms of computational cost. The numerical convergence of integrals is very slow as the system size increases and the grid resolution gets finer. Fortunately, a helpful hint for the computation of the saddle-point action $S_c(\xi)$ is available from the numerical work reported in Ref. \cite{grafke_etal}, where it is pointed out that for large negative velocity gradients and at a given noise strength $g$, $S_c(\xi)$ can be retrieved with good accuracy from the vgPDF $\rho_g(\xi)$ as
\be
S_c(\xi) \simeq - g^2\kappa(g) \ln \left [ \frac{\rho_g(\xi)}{\rho_g(0)} \right ] \ , \ \label{s_Sc}
\ee
where $\kappa(g)$ is a $g-$dependent empirical correction factor. 

It follows, now, under the light of Eq. (\ref{eff_vgpdf}), that $\kappa(g)$ is nothing more than $(g_R/g)^2$, and, therefore, it should depend on $\xi$ as well.  From such a perspective, one finds that the relevance of Eq. (\ref{s_Sc}) is fortuitously based on the fact that $c(\xi)$, as defined in (\ref{c-xi}), is in general a slowly varying function of $\xi$. As a point of pragmatic methodology, we are going to rely on Eq. (\ref{s_Sc}) as an effective way to obtain a reasonable evaluation of the saddle-point action. However, to make a clear distinction between what would be the exact saddle-point action {\it{versus}} the one approximated by (\ref{s_Sc}), we refer to the RHS of (\ref{s_Sc}) as the {\it{surrogate saddle-point action}} $S_{sc}(\xi)$.

Regarding the evaluation of the perturbative functionals $I_1 [p^c,u^c]$ and $I_2 [p^c]$, while the full numerical approach is very slowly convergent, if based on the Chernykh-Stepanov numerical solutions of Eqs. (\ref{sp1}) and (\ref{sp2}), we have found that approximate analytical expressions for $u^c(x,t)$ and $p^c(x,t)$ lead to considerable improvement by way of standard numerical integration packages. Below, we first discuss such analytical approximations and, afterwards, focus on the determination of $S_{sc}(\xi)$, $I_1 [p^c,u^c]$, and $I_2 [p^c]$.
\vspace{0.2cm}

{\leftline{\it{Analytical Approximations for the Instanton Fields}}}
\vspace{0.2cm}

In the asymptotic limit of small velocity gradients, instantons can be well approximated as the solutions of Eqs. (\ref{sp1}) and (\ref{sp2}) simplified by the suppression of nonlinear terms. Working in Fourier space, where
\bea 
&& \tilde p(k,t) \equiv \int dx \ p(x,t) \exp(-ikx) \ , \ \label{fp} \\
&& \tilde u(k,t) \equiv \int dx \ u(x,t) \exp(-ikx) \ , \ \label{fu}
\eea 
it is straightforward to find, under the linear approximation, that
\bea
&& \tilde u^c(k,t) = \lambda^c \sqrt{\frac{\pi}{2}} k \exp \left [ -k^2\left (|t|+\frac{1}{2} \right ) \right ]  \label{fuc}  \ , \ \\
&& \tilde p^c(k,t) = -i  \lambda^c k \exp( k^2 t) \Theta(-t) \equiv \tilde p^{(0)}(k,t) \ . \ \label{fpc} 
\eea
Taking $\lambda^c \equiv -i \lambda$, we get, from (\ref{fuc}), 
$\xi = \lambda /2$, the velocity gradient at $(x,t)=(0,0)$. From now on 
it is assumed, thus, that $\lambda$ is a negative real number.

Note that if we write the exact solution for the instanton response field as
\be
p^c(x,t) = p^{(0)}(x,t) + \delta p^c(x,t) \ , \ \label{pc_full}
\ee
then $\delta p^c(x,t)$ has to satisfy the boundary conditions
\be
\delta p^c (x,- \infty) = \delta p^c(x,0^-) = 0 \ , \ \label{vbc}
\ee
since, as it can be inferred from (\ref{fpc}), $p^{(0)}(x,t)$ saturates the boundary conditions for 
$p^c(x,t)$, already stated in our former discussion of Eqs. (\ref{sp1} - \ref{sp3}). 

The vanishing boundary conditions (\ref{vbc}) suggest that $\delta p^c(x,t)$ can be taken as a perturbation field, which is clearly a true fact for asymptotically small times $t$. Accordingly, the instanton velocity field can be expanded as a functional Taylor series,
\begin{widetext}
\be
u^c(x,t) = u^{(0)}(x,t) + \sum_{n=1}^\infty \int \left [ \prod_{i=1}^n dx'_i dt'_i \delta p^c(x_i',t_i') \right ] F_n(x,t, \{x',t'\}_n )  \ , \ \label{uc_full}
\ee
\end{widetext}
where 
\be
\{x',t'\}_n \equiv \{ x_1',x_2',...,x_n', t_1', t_2',...,t_n' \} 
\ee
and the many variable kernel $F_n(x,t, \{x',t'\}_n )$ is a functional of $p^{(0)}(x,t)$. Note that $u^{(0)}(x,t)$ is independent
(in the functional sense) of $\delta p^c(x,t)$. An infinite hierarchy of equations is obtained for $F_n(x,t, \{x',t'\}_n )$, when (\ref{pc_full}) and (\ref{uc_full})
are substituted into the saddle-point Eqs. (\ref{sp1}) and (\ref{sp2}). In general, $\partial_t F_n(x,t, \{x',t'\}_n )$ will depend in a nonlinear way on the set of $F_m's$, with $m \leq n$.

The interesting news here is that it is possible to get a closed analytical solution for $u^{(0)}(x,t)$. We find, in Fourier space, that $\tilde u^{(0)}(k,t)$ is the sum of two contributions,
\be
\tilde u^{(0)}(k,t) = \lambda^c \tilde F_0^{(1)} (k,t) + (\lambda^c)^2 \tilde F_0^{(2)} (k,t) \ , \ \label{u0}
\ee
where $\lambda^c \tilde F_0^{(1)}(k,t)$ is exactly the same as (\ref{fuc}),
and
\bea
&& \tilde F_0^{(2)} (k,t) = \frac{i k}{32} \sqrt{\frac{3 \pi k^2}{4}} \exp \left [ k^2 \left ( |t| + \frac{1}{2} \right ) \right ] \times \nonumber \\
&&\times \Gamma \left (-\frac{1}{2}, \frac{3 k^2}{2} \left ( |t|+ \frac{1}{2} \right ) \right ) 
-\frac{i k^3}{32} \sqrt{\frac{\pi}{3 k^2}} \times \nonumber \\
&&\times \exp \left [ k^2 \left ( |t| + \frac{1}{2} \right ) \right ] \Gamma \left (\frac{1}{2}, \frac{3 k^2}{2} \left ( |t|+ \frac{1}{2} \right ) \right ) \ , \ \nonumber \\
\label{F02}
\eea
a result expressed in terms of the incomplete Gamma function, $\Gamma(x,y) = \int_y^\infty t^{x-1} \exp(-t) dt$.
From Eq. (\ref{u0}) (using, again, $\lambda^c \equiv -i \lambda$), the velocity gradient at $(x,t)=(0,0)$ can be readily 
computed, in the approximation where $u^c(x,t) = u^{(0)}(x,t)$, as
\bea 
&& \xi \equiv \partial_x u^c(x,0)|_{x=0} = \frac{i}{2 \pi}  \int dk k \tilde ~ u^{(0)} (k,0) \nonumber \\
&&=\frac{\lambda}{2} + \frac{3 -2\sqrt{3}}{24} \lambda^2 \ , \ 
\eea
which, upon inversion leads to
\be
\lambda = 2 \frac{\sqrt{3}  - \sqrt{3 + 2(3 - 2 \sqrt{3}) \xi }}{2-\sqrt{3}} \ . \ \label{lambda-xi}
\ee
In order to see how accurate is Eq. (\ref{lambda-xi}), we have computed the numerical instantons from Eqs. (\ref{sp1}-\ref{sp3}), along the lines of the Chernykh-Stepanov procedure, implemented through the pseudo-espectral method for a system with size $200$ (recall that $L=1$), and $2^{10}$ Fourier modes. The time evolution is realized in the frame of a second order Adams-Bashfort time-difference scheme with time step $\delta t = 10/2048 \simeq 5 \times 10^{-3}$ and total integration time $T = 200$. Since instantons evolve within the typical integral time scale $T_0 \sim 1/|\lambda|$, we have investigated the range $0.5 \leq |\lambda| \leq 20.0$, so that $ \delta t \ll T_0 \ll T$.

\begin{figure}[ht]
\hspace{0.1cm} \includegraphics[width=.5\textwidth]{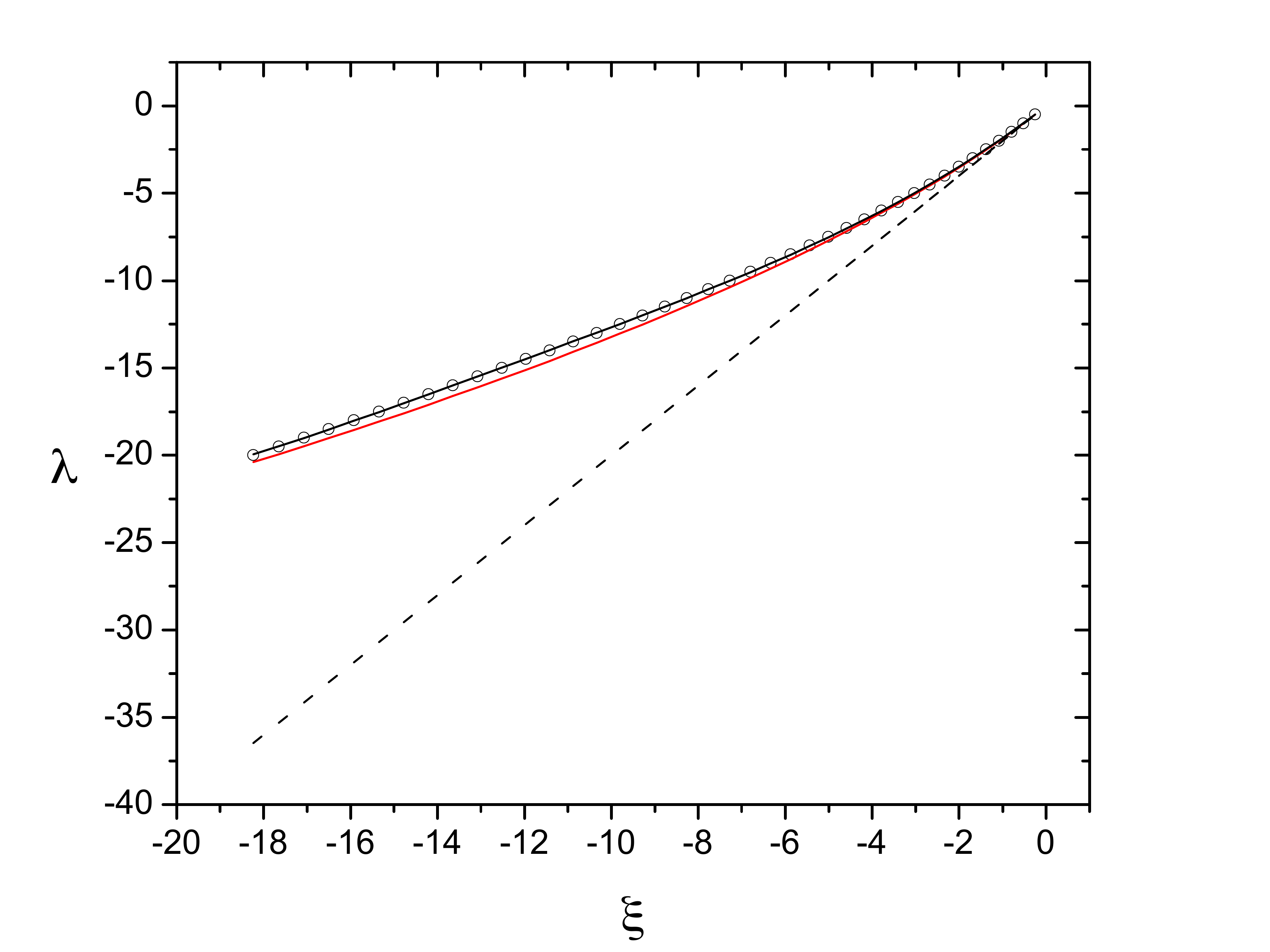}
\caption{The Lagrange multiplier $\lambda$ is given as a function of the velocity gradient $\xi$. Open circles represent values obtained from the numerical solutions of Eqs. (\ref{sp1}-\ref{sp3}) (the black solid line is just a polynomial interpolation of the numerical data); red solid line: approximated instanton relation, Eq. (\ref{lambda-xi}); dashed line: $\lambda = 2 \xi$, which holds for asymptotically small velocity gradients.
}
\label{lambda_fig}
\end{figure}

As we can see from Fig. \ref{lambda_fig}, the comparison between the predicted relation (\ref{lambda-xi}) and the one obtained from the numerical instantons is reasonably accurate.

\vspace{0.2cm}

{\leftline{\it{The Surrogate Saddle-Point Action}}}
\vspace{0.2cm}

While we expect that the approximate instanton fields given by Eqs. (\ref{fpc}) and (\ref{u0}) can be useful for the evaluation of $I_1 [p^c,u^c]$ and $I_2 [p^c]$, up to lowest non-trivial order in the functional perturbative expansion around $p^{(0)}(x,t)$, they are, unfortunately, unable to provide the observed dependence of the nonperturbative MSRJD action $S_c(\xi)$ with the velocity gradient $\xi$. In fact, $p^{(0)}(x,t)$ is proportional to $\lambda$, leading, from (\ref{sp_action}), to $S_c(\xi) = \lambda^2/4$, a result that is not supported by Eq. (\ref{s_Sc}) with the input of numerical vgPDFs \cite{grafke_etal}.

Taking advantage of the results reported in Ref. \cite{grafke_etal} for the case of noise strength parameter $g=1.7$, a flow regime close to the onset of intermittency, we set $\kappa(g) = (0.92)^2$ and write down the surrogate saddle-point action (\ref{s_Sc}) as
\be
S_{sc}(\xi) \simeq - (0.92 \times 1.7)^2 \ln \left [ \frac{\rho_{1.7}(\xi)}{\rho_{1.7}(0)} \right ] \ . \ \label{s_sc2}
\ee
We have carried out direct numerical simulations to obtain the surrogate action (\ref{s_sc2}) and a set of vgPDFs for other values of $g$, with the purpose of checking  (\ref{vgpdf2}) in the approximation given by (\ref{c-expansion}). 

The stochastic Burgers equation is solved with a fully dealised pseudo-spectral method in $N=2048$ collocation points \cite{canuto_etal} by employing a $2^\text{nd}$ order predictor-corrector time marching scheme \cite{kloeden_etal}. As in our numerical solution of the instanton fields, the domain size is taken to be $200L$. Velocity gradients are saved every 30 time steps after a suitable transient time, during a total simulation time $T \approx 1.2\times 10^7$.

 \begin{figure}[ht]
\includegraphics[width=.5\textwidth]{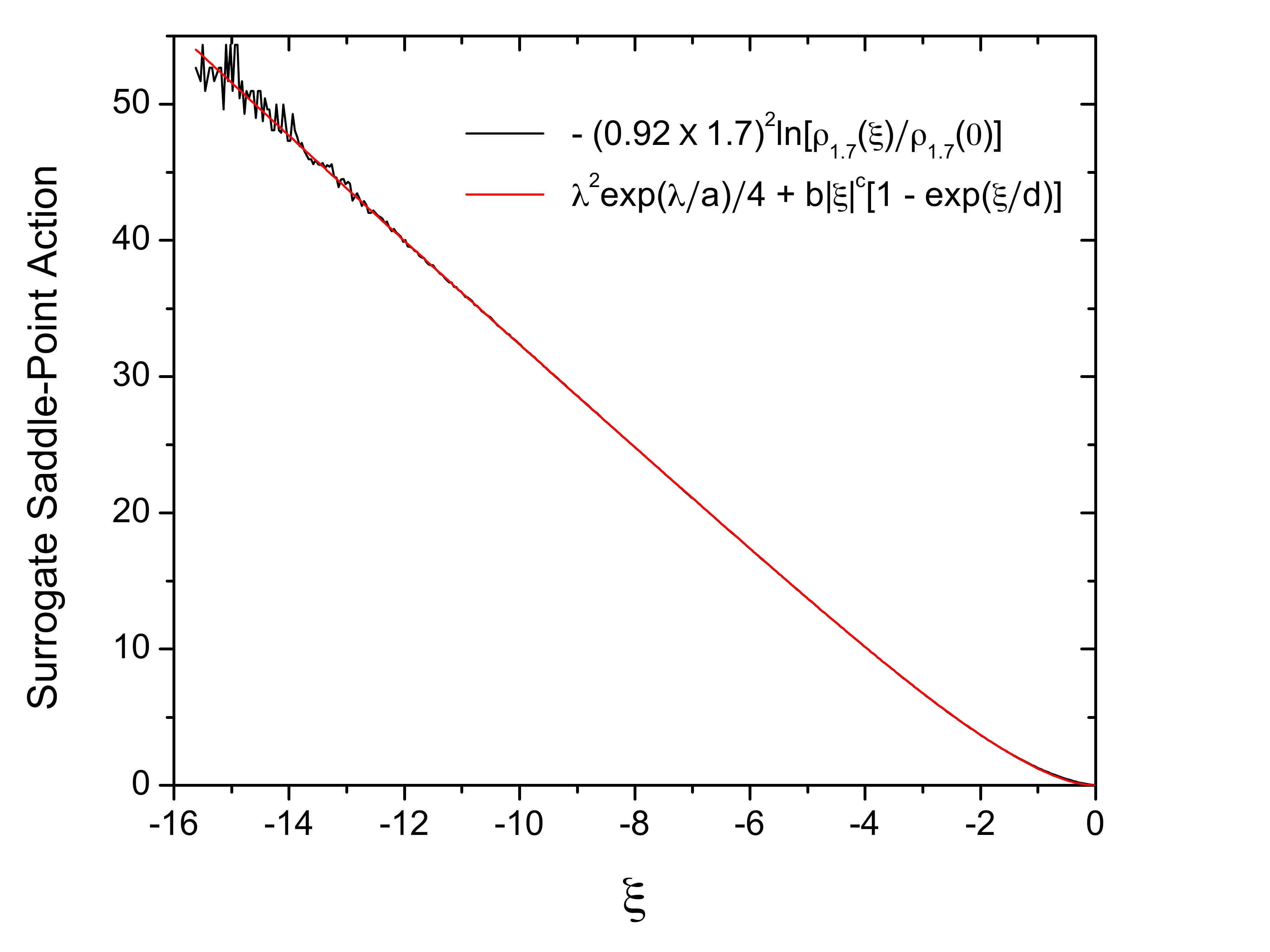}
\caption{Comparison between the surrogate saddle-point action, as prescribed by Grafke {\it{et al}}. \cite{grafke_etal} for the case of noise strength parameter $g=1.7$ (black solid line), and a four-parameter fitting function (red line) which provides distinct power law asymptotics for domains of small and large velocity gradients.} \label{surrogate_action}
\end{figure}

A useful and accurate fitting of the surrogate saddle-point action (\ref{s_sc2}) can be defined as
\be
S_{sc}(\xi) = \frac{\lambda^2}{4} \exp \left ( \frac{\lambda}{a} \right ) + b|\xi|^c \left [1 - \exp \left ( \frac{\xi}{d} \right  ) \right ] \ , \ \label{interpol}
\ee
where $\lambda$ is given by (\ref{lambda-xi}), and $a=2.046$, $b=2.407$, $c=1.132$, and $d=2.195$ are optimal fitting parameters. The result is shown in Fig. \ref{surrogate_action}. 

The interpolation (\ref{interpol}) is actually consistent with the behavior of the local stretching exponent for the saddle-point action, which shows a quick drop from $\theta(\xi) = 2$ at small velocity gradients to $\theta(\xi) \simeq 1.16$ as $|\xi|$ grows, a fact verified from direct numerical simulations of the Burgers equation as well \cite{grafke_etal, gotoh-kraichnan}. The main benefit of using (\ref{interpol}) instead of the raw surrogate saddle-point action derived from $\rho_{1.7}(\xi)$ is that it yields a smooth interpolation of data, circumventing error fluctuations that grow at larger values of $|\xi|$.
\vspace{0.2cm}

{\leftline{\it{Evaluation of $I_1 [p^c,u^c]$ and $I_2 [p^c]$}}}
\vspace{0.2cm}

 \begin{figure}[t]
\includegraphics[width=.6\textwidth]{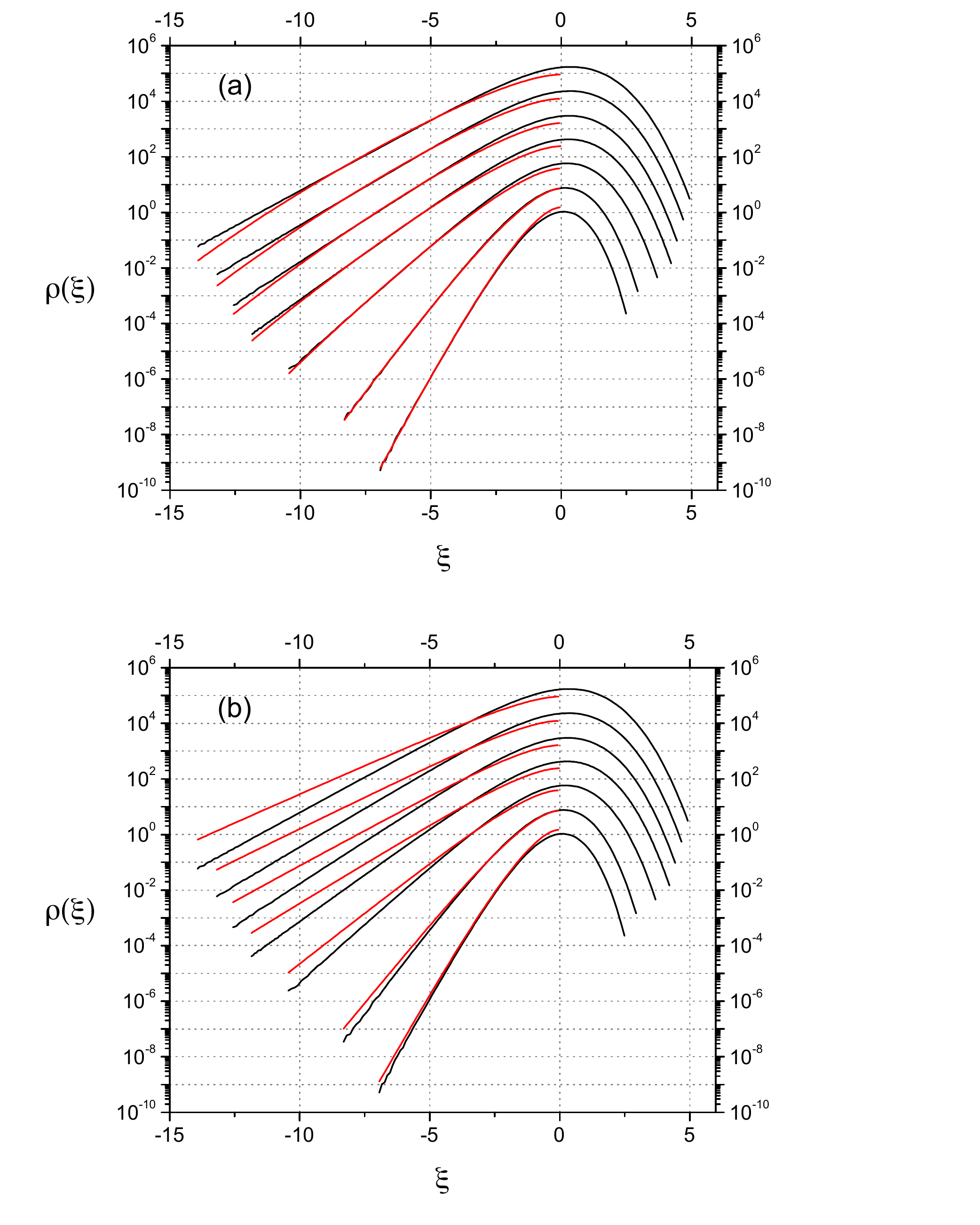}
\caption{Modeled (red lines) and empirical (black lines) vgPDFs are compared for noise strengths $g=1.0, 1.2, 1.5, 1.7, 1.8, 1.9$, and $2.0$. They have been shifted along the vertical axis to ease visualization, and their associated values of $g$ grow from the bottom to the top in each one of the PDF sets. Figures (a) and (b) give the modeled vgPDFs that include and neglect, respectively, the effects of fluctuations around instantons.} \label{vgPDFs}
\end{figure}

\begin{figure}[]
\includegraphics[width=.49\textwidth]{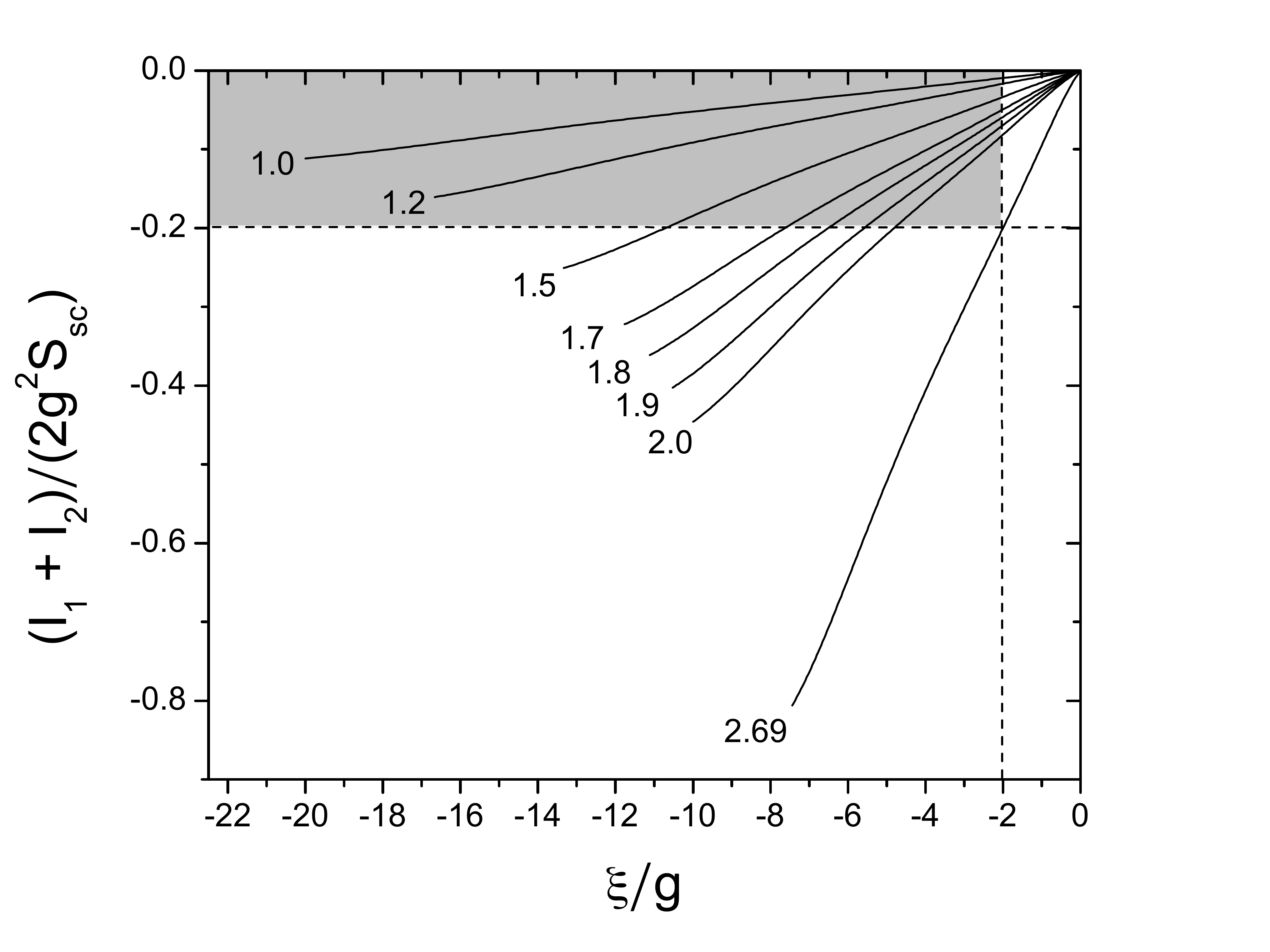}
\caption{Solid lines, labeled by values of $g$, represent relative corrections to the MSRJD surrogate saddle-point action due to fluctuations around instantons. The intersection points of each one of the solid lines with the vertical and horizontal dashed lines define the range of normalized velocity gradients $\xi/g$ where the perturbative cumulant expansion is assumed to work (highlighted region in the plot).} \label{pert_domain}
\end{figure}

Since $I_1 [p^c,u^c]$ is a linear functional of $u^c(x,t)$ we can write, from (\ref{pc_full}) and (\ref{u0}), that
\bea 
&&I_1 [p^c,u^c] + I_2 [p^c] = I_1 [p^{(0)},\lambda^c F_0^{(1)}]  + \nonumber \\
&&+ I_1 [p^{(0)}, (\lambda^c)^2 F_0^{(2)}] + I_2 [p^{(0)}] + \mathcal{O}[\delta p^c] \nonumber \ . \ \label{I1I2dp} \\
\eea
In order to evaluate the first three terms on the RHS of (\ref{I1I2dp}), it is interesting, for the sake of fast numerical convergence, 
to write the two-point correlation functions (\ref{Gpu}) and (\ref{Guu}) in Fourier space, viz.,
\bea 
&&\tilde G_{pu} (k,t,t') = \int d x \ G_{pu}(x,0,t,t') \exp(-ikx) = \nonumber \\
&&= -i g^2 \exp \left [ - (t' -t)k^2 \right ] \Theta(t'-t) \ , \ \label{fGpu} \\
&&\tilde G_{uu} (k,t,t') = \int d x \ G_{uu}(x,0,t,t') \exp(-ikx) = \nonumber \\
&&= g^2 \sqrt{\frac{\pi}{2}} \exp \left 
[ - \left (|t' -t| + \frac{1}{2} \right ) k^2 \right ] \ . \ \nonumber \label{fGuu} \\
\eea 
We have, from (\ref{I1}), (\ref{I2}), (\ref{fGpu}), and (\ref{fGuu}),
\begin{widetext}
\bea 
&&I_1[p^{(0)},\lambda^c F_0^{(1)}]  = \frac{\lambda^c }{2 \pi^2} \int_{t,t'<0} dt dt' \int dk dk' ~  k(k+k') \tilde p^{(0)}(k,t) \tilde F_0^{(1)} (-k,t')  
\tilde G_{uu} (k',t,t')   \tilde G_{pu}(k + k',t',t) \nonumber \\
&& = \frac{\lambda^2 g^4}{8 \pi} \int dk dk' ~  \frac{k(k+k')}{k^2+k'^2+(k+k')^2}  \exp \left [ -\frac{1}{2} \left (k^2+k'^2 \right ) \right ]  \ , \ \label{I1n}
\\
&&I_2[p^{(0)}] = - \frac{1}{2(2 \pi)^2} \int_{t,t'<0} dt dt' \int dk dk'  k^2 \tilde p^{(0)}(k,t) \tilde p^{(0)}(-k,t')  
\tilde G_{uu} (k',t,t')   \tilde G_{uu}(k + k',t,t') \nonumber \\
&& = -\frac{\lambda^2 g^4}{16 \pi} \int dk dk' \frac{k^2}{k^2+k'^2+(k+k')^2} \exp \left [ -\frac{1}{2} \left (k'^2 + (k+k')^2 \right ) \right ] \ , \ \label{I2n}
\eea
\end{widetext}
implying that 
\be
I_1[p^{(0)}, \lambda^c F_0^{(1)}] = - I_2[p^{(0)}] = (3-\sqrt{3}) \lambda^2 g^4/24
\ee
and, according to (\ref{I1I2dp}),
\be
I_1 [p^c,u^c] + I_2 [p^c] = I_1 [p^{(0)},(\lambda^c)^2 F_0^{(2)}]  + \mathcal{O}[\delta p^c] \ . \ \label{I1dp}
\ee
A straightforward numerical evaluation yields, from (\ref{I1}),
\be
I_1 [p^{(0)},(\lambda^c)^2 F_0^{(2)}] \simeq 1.6 \times 10^{-3} \lambda^3 g^4 \ . \
\ee
Eqs. (\ref{vgpdf2}), (\ref{c-expansion}), (\ref{S1squared}), and (\ref{I1dp}) provide all the necessary ingredients we need to put forward an improved expression for the vgPDF
tails, more concretely,
\be
\rho_g(\xi) = C(g) \exp \left [- \frac{1}{g^2} S_{sc}(\xi) + \frac{1}{2g^4} I_1 [p^{(0)}, (\lambda^c)^2 F_0^{(2)}] \right ] \ , \ \label{rho_complete}
\ee
where $C(g)$ is a normalization constant that cannot be determined from the instanton approach, since it depends on the detailed shape of the vgPDF for $- \infty < \xi < \infty$, while (\ref{rho_complete}) refers, in principle, to negative velocity gradients which are some standard deviations away from the mean. The relevance of the saddle-point computational strategy (including fluctuations), however, can be assessed from adjustments of $C(g)$ that produce the best matches between the predicted vgPDFs, Eq. (\ref{rho_complete}), and the empirical ones, obtained from the direct numerical simulations of the stochastic Burgers equation \cite{grafke_etal}. We do exactly so, using the least squares method, in the velocity gradient range $-5g \leq \xi \leq -3g$.

Comparisons between the predicted and empirical vgPDFs are shown in Fig. \ref{vgPDFs}, for $g=1.0$, $1.2$, $1.5$, $1.7$, $1.8$, $1.9$, and $2.0$, with and without the fluctuation correction term proportional to $I_1 [p^{(0)},(\lambda^c)^2 F_0^{(2)}]$, as it appears in (\ref{rho_complete}).

We find that the surrogate saddle-point action is in fact a very good approximation to the exact one, by inspecting the vgPDF for $g=1.0$, when the cumulant contribution is almost negligible. As $g$ grows, the relative cumulant contributions grow as well, and become essential in order to attain accurate modeling of vgPDF tails. For $g=1.7$, as an example, we clearly verify the existence of a fat left tail, and an excellent agreement between modeled and empirical vgPDFs that extends for about four decades.

As it can be seen from Fig. \ref{vgPDFs}, as $g$ grows, the velocity gradient regions where the agreement between the predicted and the empirical vgPDFs is reasonably good shrink in size. This is, of course, expected under general lines, since the cumulant expansion is a perturbative method supposed to break down when the amplitude of saddle-point configurations become large enough, which in our particular case takes place for large negative velocity gradients.

\section{Perturbative Domain}

We find, from an analysis of the vgPDFs depicted in Fig. \ref{vgPDFs}, that a fine matching between the predicted and the empirical vgPDFs holds for $|\xi| > 2g$, but starts to lose accuracy when velocity gradients are such that the second order cumulant expansion contributions, $(I_1[p^c,u^c]+I_2[p^c])/2g^4$, are of the order of $20 \%$ (in absolute value) of the dominant saddle-point contributions, $S_{sc}(\xi)/g^2$. We report, in Fig. \ref{pert_domain}, how the ratio between these two quantities depends on the velocity gradient $\xi$ for the several investigated values of the noise strength parameter $g$, up to $g=2.0$. It can be estimated in this way, then, that $g \simeq 2.7$ is an upper bound for the usefulness of the cumulant expansion method.

\section{Conclusions}

Notwithstanding the fact that the instanton approach to Burgers intermittency was introduced around two decades ago \cite{gura-migdal,balko_etal}, the modeling of its preasymptotic, but already fat-tailed, vgPDFs has been a persistent puzzle along the years. The central issue underlying such a difficulty is that instantons are supposed to yield an asymptotic description of far vgPDF tails, which are not accessible, in general, from direct numerical simulations.

Previous results, derived in the context of Lagrangian turbulence \cite{mori2,apol}, have indicated that non-Gaussian fluctuations of important fluid dynamic observables, such as velocity gradients, can be perturbatively investigated at the onset of intermittency by means of the cumulant expansion technique. The main lesson taken from these studies is that at the onset of intermittency, the MSRJD saddle-point action gets its heat-kernel and noise correlator function renormalized as a dynamical effect of fluctuations around instantons. In this way, accurate comparisons between analytical and empirical vgPDFs have been achieved.

Inspired by such ideas, we have applied a similar approach to the problem of stochastic Burgers hydrodynamics, which is able to predict the detailed shape of vgPDF left tails at the onset of intermittency. Our results \textcolor{black}{show that an account of fluctuations around instantons is in fact necessary to render the instanton approach a meaningful tool for the modeling of Burgers intermittency, as emphasized by Grafke {\it{et al}}. \cite{grafke_etal}.}

It is likely that the field theoretical treatment addressed in this work can be extended to other related problems, like the transport of passive scalars \cite{balko_etal} and the statistics of vorticity in three or two-dimensional turbulence \cite{mori_vort, smith_yakhot, falko_lebed}.

Moving forward to the study of vgPDF tails for fully developed turbulent regimes, far beyond the onset of intermittency, is another challenging task. The cumulant expansion method breaks down and improved techniques for evaluating the path-integration over fluctuations around the instantons are in order, ultimately related to the analysis of functional Hessian determinants \cite{rajaraman,forman,mckane-tarlie,muratore}. However, it is not clear at all if alternative path-integration methods will be of any relevance without the consideration of further improved analytical approximations for the instanton solutions. \textcolor{black}{Also, as a point to be clarified in further studies, one may wonder if Gaussian fluctuations are indeed enough {\it{per se}} to model in a satisfactory way the whole extension of vgPDF tails, since an analogous approach is known to lead to inconsistencies in the multifractal description of intermittency \cite{frisch_etal}.}


\acknowledgments

This work has been partially supported by CAPES, CNPq, and 
FACEPE. We thank T. Grafke for kind email exchange about the work 
of Ref. \cite{grafke_etal}.


\begin{references}
\bibitem{comment1} The velocity gradient variance is $g^2/2$ in the statistically stationary regime. To prove it, just 
multiply both sides of Eq. (\ref{burgers_eq}) by $u(x,t)$ and take, using the It\^o's Lemma, the average of the resulting equation.
\end{references}
\end{document}